# Add drop multiplexers for terahertz communications using two-wire waveguide based plasmonic circuits

Yang Cao, Kathirvel Nallappan, Guofu Xu, and Maksim Skorobogatiy[*]

*Department of Engineering Physics, École Polytechnique de Montréal, Montreal, Québec, H3T 1J4, Canada*
*maksim.skorobogatiy@polymtl.ca*

**Abstract:** Terahertz (THz) band is considered as the next frontier in wireless communications. The emerging THz multiplexing techniques are expected to dramatically increase the information capacity of THz communications far beyond a single channel limit. In this work, we explore the THz frequency-division multiplexing modality enabled by novel add-drop multiplexer (ADM) design. Based on modular two-wire plasmonic waveguides fabricated using additive manufacturing and metallization techniques, we demonstrate four-port THz ADMs containing grating-loaded side couplers for operation at ~140 GHz carrier frequency. Particular attention is payed to the design of plasmonic waveguide Bragg gratings and directional couplers capable of splitting broadband THz light in spectral and spatial domains, respectively. Finally, we demonstrate multiplexing and demultiplexing of THz signals with bit rates up to 6 Gbps using the developed ADMs. We believe that proposed plasmonic circuits hold strong potential to provide robust integrated solutions for analogue signal processing in the upcoming THz communications.

## 1. Introduction

The global mobile data traffic is expected to reach 77.49 exabytes per month by the year 2022 [1]. According to Shannon-Hartley theorem [2], in the presence of noise, the maximal bit rate supported by the transmission channel is proportional to the available bandwidth. Currently, most of the wireless systems operate in the overcrowded microwave band which is not sufficient to meet the bandwidth demand of the near future. Shifting the carrier wave towards higher frequencies is inevitable in order to accommodate the imminent surge in data volume [3,4]. Therefore, the terahertz (THz) frequency band (0.1-10 THz) is considered as the next frontier in wireless communications [5,6]. To date, there are already several demonstrations of free-space ultra-high bit rate data transmission (> 100 Gbps) using a single-channel THz link having carrier frequency in 0.3-0.4 THz together with various optical multiplexing (ex. optical frequency and polarization division multiplexing) and modulation techniques (ex. quadrature amplitude modulation and quadrature phase-shift keying) [7-9]. The capacity of the THz communication system can be further increased by using various THz multiplexing techniques [10-12]. Among them, the frequency-division multiplexing (FDM), in which several discrete carrier frequencies support distinct users, is routinely used in fiber optic communications to multiply the data-throughput capacity. In optical networks, one typically uses carrier waves in C-band (infrared) with channel spacing less than 10 GHz, thus allowing simultaneous transmission of hundreds of channels with hundred MHz bandwidth each [13]. In THz communications, the current studies place the carrier wave frequency in 100 – 300 GHz range with a single channel bandwidth of ~10 – 40 GHz, which can allow spectral allocation and simultaneous utilisation of several tens of communication channels within this spectral range.

The key component in designing FDM systems is the add-drop multiplexer (ADM). It allows multiple carrier frequencies to share the same frontend, thus reducing the overall size and complexity of the communication networks [14,15]. In optical FDM networks, the add-drop multiplexing is usually realized by the combination of mature optical elements such as ring resonator, phase-shifted grating, Bragg grating, arrayed waveguide grating, circulator, and coupler on the platform of silica-based planar lightwave circuits and fiber arrays [16-19]. However, the ADMs capable of operation at THz frequencies are still in the early development due to lack of universal standards for the THz communication components, as well as relative abundance of possible, while untested, fabrications routes.

Using the directional coupling between two adjacent waveguides or the leaky wave of a single deformed waveguide, THz ADMs have been proposed based on parallel-plate waveguides and silicon substrates [15, 20-24]. Particularly, the silicon substrates also have great potential in developing several feasible THz ADM designs, such as diplexer and bandpass filters [25], ring resonators [26], and topological insulators [27]. Nevertheless, add/drop frequency tunability is problematic, and the top-of-the-line expensive fab infrastructure is required to fabricate silicon-based circuits. Besides, a THz ADM prototype composed of a lossy Y-coupler and an external low-reflectivity Bragg grating was also fabricated using two-wire plasmonic waveguides [28,29]. However, most of the reported THz ADMs contain only three ports, while to implement both channel dropping and adding of THz carrier waves, cascaded ADMs or four port ADMs have to be employed in the communication networks.

In what follows, we report the design and characterization of four-port THz ADMs for FDM communications. The core of our system is modular micro-encapsulated two-wire plasmonic waveguides fabricated using stereolithography (SLA) 3D printing and wet chemistry metal deposition techniques reported recently in [28]. Due to low-cost and wide availability of the fabrication infrastructure, the two-wire photonic circuits in which the plastic surfaces are selectively metallized are highly suitable for rapid prototyping and mass production. Such waveguide features low transmission and bending loss, low group velocity dispersion (GVD), broadband operation, as well as high coupling coefficient between connected waveguide sections and with the conventional linearly polarized THz sources [30,31]. Additionally, this two-wire waveguide provides a versatile platform to build highly reconfigurable and convenient to handle THz circuits, including the desired ADMs that enable advanced signal processing functionalities in THz communications. By replacing the decisive modular element of ADM circuits, the drop/add frequencies and the corresponding channel bandwidths can be readily changed without affecting the rest of circuits. Furthermore, thermal and mechanical tuning of the proposed circuits is readily achievable as modal fields of air-core two-wire waveguides are easily accessible. This opens an important opportunity for such structures toward meeting the requirement of dynamic band allocation in THz communication networks [32].

One key element of our system is a waveguide Bragg grating (WBG) that serves as a wavelength-selective mirror [33]. By integrating the WBG into optical circuits, a tunable stopband and passband in the transmission and reflection spectra can be achieved. Hindered by the lack of THz circulators compatible with circuits [34,35], the directional coupler is used as an alternative diplexer to combine THz signals in spatial domain, which is another key element of the proposed ADM circuit. Particularly, we develop an interferometric THz ADM where uniform Bragg grating is symmetrically added on two side coupled two-wire waveguides, which is a concept adapted from the mid-infrared range [36-38]. Compared with other ADM designs of similar configurations, such as grating-assisted contra-directional coupler [39] and grating-loaded Mach-Zehnder interferometer [40], the aforementioned design has advantages of compact structure, simple design criteria, single-mode operation, while allowing relatively relaxed fabrication tolerances. Additionally, due to the low reflectance at resonant frequency of a distributed feedback reflector (quarter wave defect) inserted into the proposed two-wire WBG, another ADM design that is universal in mid-infrared band where phase-shifted gratings are placed on directional coupler sections is not a promising alternative for our circuit [19,41].

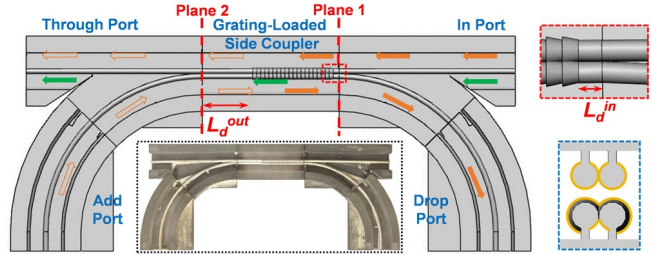

Fig. 1 Schematic of the proposed four-port ADM circuit (top part is not shown) comprising a grating-loaded side coupler placed between two Y splitters. $L_d^{in}$ and $L_d^{out}$ are the distances between the grating end-facets and the side coupler edges (see an Insert in the red dotted box). The orange arrows indicate the path of a THz carrier wave that falls within the grating stopband (dropped or added signals). The green arrows indicate the path of a THz carrier wave outside of the grating stopband (through signal). The arrows of solid colors correspond to the waves launched into the In port and then either dropped or guided through. The hollow arrows correspond to the waves launched into the Add port and then added to Through port. Insert in the blue dotted box presents a cross section of the grating-loaded side coupler. Insert in the black dotted box presents an experimental four-port ADM circuit (top part is not shown).

This paper is organized as follows. Firstly, we develop high-reflectivity WBGs based on the two-wire THz plasmonic waveguides, with one of the wires featuring a periodic array of end-to-end connected truncated cones. As a result, we experimentally demonstrate stopbands of up to 14 GHz with a center wavelength that can be positioned anywhere in the 120-160 GHz range. Secondly, we design and fabricate directional couplers based on two adjacent two-wire waveguides, and then confirm experimentally that a several centimeter-long coupler is capable of full-cycle coupling at 140 GHz carrier frequency over a broad bandwidth exceeding 60 GHz. Finally, we propose a four-port THz ADM circuit, where the grating-loaded side coupler plays the crucial role which allows to spatially separate (drop) and add the THz wavelengths that fall within the grating stopband, while letting all the other wavelengths within the coupler bandwidth to pass through. Two wideband Y splitters connected with the coupler are used to guide THz signals to/from the grating-loaded side coupler section from/to the desired ports of ADM circuits (see Fig. 1). Channel dropping and adding functionality of ADMs for several THz carrier waves in the vicinity of ~140 GHz modulated up to 6 Gbps were demonstrated in experiments.

## 2. Results

**Two-wire waveguide-based Bragg grating**

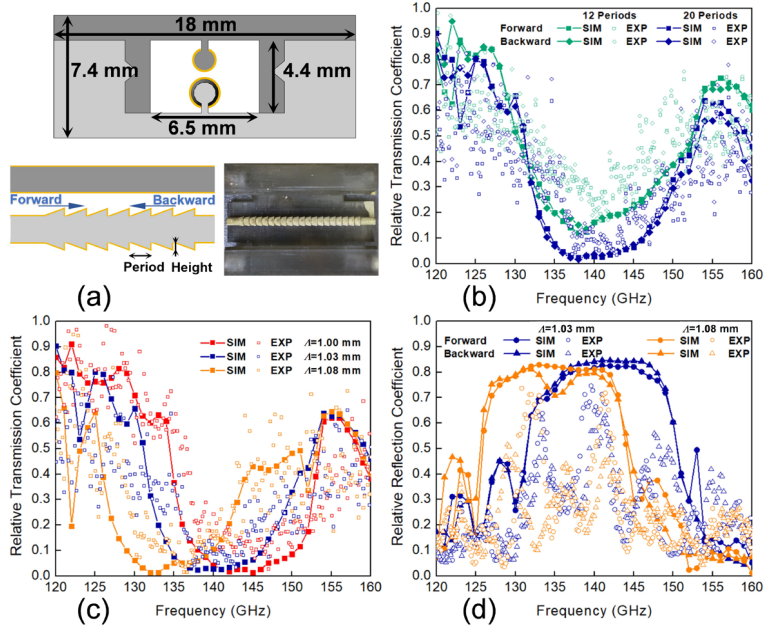

Fig. 2 Numerical and experimental studies on the two-wire WBGs. (a) Schematic of the cross section of a two-wire WBG comprising two 3D-printed parts (different shades of gray), its side view, as well as the photograph of a fabricated WBG (top half (dark gray) removed). The gratings featuring a periodic sequence of truncated cones (light gray) are suspended in air (white) on dielectric supports and encapsulated within a plastic cage (light and dark gray). Both the cylindrical wire (dark gray) and the gratings are covered with a silver layer (gold) by wet chemistry deposition. (b) The numerical and experimental relative transmission coefficients of 2.5cm-long WBG sections comprising 12 and 20 periods ($\Lambda$=1.03 mm, $H$=0.21 mm) with two possible travelling directions of the THz light. (c) The numerical and experimental relative transmission coefficients of 2.5cm-long WBGs containing 20 periods ($H$=0.21 mm) as a function of the period length $\Lambda$. The THz light travels in the forward direction along WBG (see Fig. 2(a)). (d) The numerical and experimental relative reflection coefficients of 2.5cm-long WBG sections containing 20 periods ($H$=0.21 mm) for two different grating period lengths, and two possible propagation directions along WBGs.

The two-wire WBGs studied in this work are introduced by periodically varying the cross section of a single wire inside a micro-encapsulated two-wire waveguide [28]. The choice of end-to-end connected truncated cones in the grating structure (see Fig. 2(a)) was experimentally found to be the most reliable and stable for printing compared to other alternative designs (see Supplementary Note 1 for details on the design and fabrication of two-wire WBGs). Such WBGs typically feature high grating strengths and wide stopbands $\Delta\lambda \sim 0.1 - 0.2 \cdot \lambda_{Bragg}$ as their strong geometrical overlap with the modal field confined in the gap between two wires. Spectral position of the grating stopband can be estimated using standard quarter wave condition $\lambda_{Bragg} \approx 2\Lambda n_{eff}$, where effective refractive index of the grating mode for two-wire waveguides is close to that of air $n_{eff} \sim 1$-1.1.

As we will see in what follows, operational bandwidth of the ADM is determined by that of the WBGs used in its design. In this work, by comparing optical properties of WBGs we find the optimal ridge height of truncated cones to be $H$=0.21 mm to ensure large operational bandwidth, reproducible optical performance, and manageable loss of ADMs. With the period length of a 20-periods WBG set to $\Lambda$=1.03 mm, the resultant stopband center frequency is 140 GHz with the corresponding bandwidth of ~18 GHz (full width at half maximum (FWHM)) in numerical simulations.

Fig. 2(b) compares the numerical and experimental results for the relative transmission coefficient of two-wire WBGs containing different numbers of periods, and two possible directions of the THz light propagation. While more periods result in deeper transmission dips, experimental WBGs show somewhat shallower transmission spectra comparted to the numerical simulations, which is attributed to imperfections in the geometrical structure of 3D printed gratings. Nevertheless, the experimental WBG with 20 periods exhibits a pronounced transmission dip near 140 GHz featuring a bandwidth of ~14 GHz (FWHM) and a minimal transmittance (by power) of less than 1%. Additionally, WBG transmission spectra are virtually symmetric when switching the input and output ports despite the WBG nonsymmetric structure. Furthermore, we confirm designability of the WBG spectral response by varying grating periodicity (see Fig. 2(c)). Reducing the grating period from 1.08 mm to 1 mm shifts the center frequency of WBG stopband from 134 GHz to 145 GHz, while measured WBG bandwidth remains largely unaffected (~14 GHz).

The reflection spectra of WBGs were experimentally characterized using a continuous-wave (CW) THz spectroscopy system together with a two-wire waveguide-based Y-coupler. Although there is a qualitative agreement between the measurements and theoretical predictions with spectral reflection peaks located within the WBG stopband, we also note strong ripples that are presented in the experimental reflection spectra (see Fig. 2(d)). These result from the standing waves formed in the cavity of the CW-THz spectroscopy setup, as well as inside of a photomixer silicon lens. Although such spectral oscillations can be somewhat mitigated via normalization with respect to a reference, they are, nevertheless, notoriously difficult to eliminate. This is especially true when the effective cavity length of a spectrometer changes during experiments, which is the case for measurements in the reflection mode when using different WBGs [42,43]. Even though, we can still see that the relative reflection coefficient of a WBG with 20 periods and $\Lambda$= 1.03 mm is ~0.7 in the spectral range between 133 GHz and 147 GHz.

# Two-wire waveguide-based directional coupler circuit

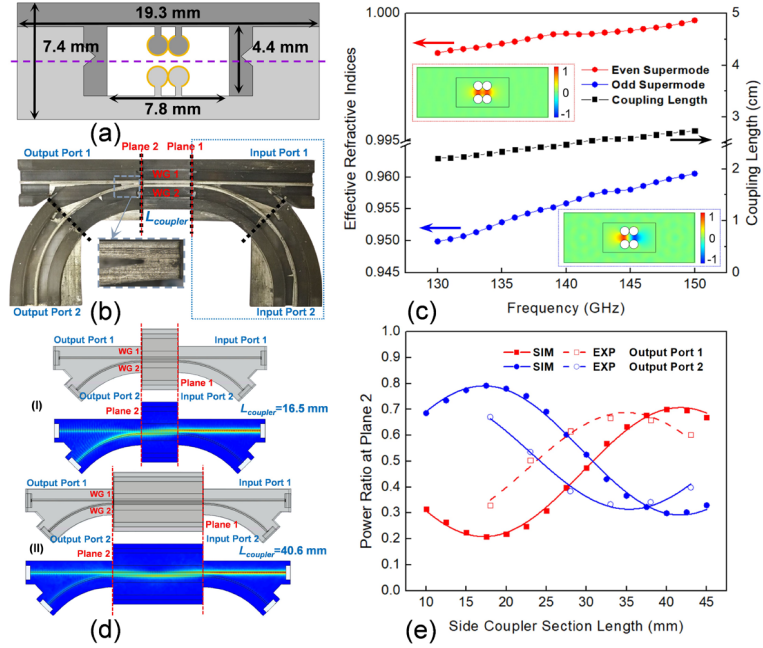

Fig. 3 Numerical and experimental studies on two-wire directional coupler circuit. (a) Schematic of the cross section of a straight side coupler featuring two touching two-wire waveguides that are suspended in air on thin dielectric supports and encapsulated within a plastic cage. Gold defines the metallic layer, light and dark gray correspond to the two halves of a device printed separately in photosensitive resin. (b) A photo of the assembled directional coupler circuit (top half removed) comprising a side coupler section placed between two Y splitters. The black dashed lines define interfaces between distinct modular sections of the circuit. Insert: the enlarged view of the Y splitter end where curved and straight waveguides meet. (c) The numerical effective refractive indices of the odd and even supermodes that are supported by the side coupler section, as well as the corresponding coupling length. Inserts: the X-component of electric field distributions for the even and odd supermodes, which is on the direction that is same with the polarization direction of the supported modes. (d) 3D models of the directional coupler circuits, and the computed electric field distributions in the circuits when using the fundamental mode of IP 1 at 140 GHz as the excitation condition. (I) Case of the half-cycle coupler $L_{coupler}$=16.5 mm that results in the maximal power transfer from Waveguide 1 to Waveguide 2. (II) Case of the full-cycle coupler $L_{coupler}$=40.6 mm that results in the maximal power transfer from Waveguide 1 to Waveguide 2 and then back to Waveguide 1. (e) The numerical derived and experimentally measured power transmittance ratio of the OP 1 and OP 2 at Plane 2 as a function of the side coupler section length $L_{coupler}$.

Experimental realization of directional coupling using free-standing THz waveguides is typically challenging. This is due to the need of precise alignment of such waveguides to maintain a fixed inter-waveguide separation (normally on a sub-mm scale), which necessitates the use of cumbersome holders, spacers, etc. [44-47]. In contrast, using high-definition 3D printing allows fabrication of the monolithic solutions that integrate waveguides, holders, and enclosure in a single step with precise control over the inter-waveguide separation. In the case of micro-encapsulated two-wire waveguides, the field of the fundamental mode is predominantly confined in the air gap between two wires [28]. Therefore, directional coupling between two separate waveguides is weak. To enhance the inter-waveguide coupling, we placed the cylindrical wires belonging to two different two-wire waveguides to touch each other and increased the enclosure width (by a single wire diameter) to accommodate these two waveguides (see

Figs. 3(a) and 3(b)).

The effective refractive indices ($n_e$ and $n_o$) of the even and odd supermodes supported by the side coupler are computed using 2D mode solver tool of COMSOL Multiphysics, while the corresponding coupling length $L_c$ between the two modes is estimated as $L_c = \lambda/2(n_e-n_o)$ (see Fig. 3(c)). For the presented design, the coupling length is $L_c \sim 24.5$ mm at ~140 GHz operational frequency, while even smaller coupling lengths (stronger coupling strengths) are possible to realize by allowing the two wires to partially overlap. To spatially separate the THz light at the output end (Plane 2) and allow a two-port input (Plane 1), two Y splitters were added at both ends of the side coupler element (see Fig. 3(b)). This configuration was then used to build the four-port ADMs. The Y splitter features a curved two-wire waveguide gradually approaching a straight one, while joining in a wedge configuration at Plane 2 (see the blue dotted region in Fig. 3(b)).

Optical performance of the complete directional coupler circuit as a function of the side coupler length (distance between Planes 1 and 2) was carried out using 3D module with ports of COMSOL Multiphysics. Particularly, we are interested in finding the relative signal intensities at Output ports 1 and 2 (OP 1,2) of the directional coupler circuit containing side couplers featuring different lengths when assuming that all the power is launched into Input port 1 (IP 1). An immediate complication is that two branches of the Y splitter used at the output side of side coupler have significantly different losses as one of them is curved and longer compared to the straight one. Therefore, instead of simply plotting the power transmittance in the OP 1 and OP 2 assuming excitation at the IP 1 expressed via the corresponding elements of the scattering matrix ($T_{op\ 1,2} = |S_{OP\ 1,2;\ IP\ 1}|^2$) (see Supplementary Note 2), we characterize the performance of side coupler section by estimating the ratio of power carried by the two waveguides (Waveguide 1 and Waveguide 2) at its end (Plane 2) using,

$$T_{CW\ 1,2} = |S_{OP\ 1,2;\ IP\ 1}|^2 / T_{Y-branch\ 1,2} \tag{1}$$

$$Ratio_{CW\ 1,2} = \frac{T_{CW\ 1,2}}{T_{CW\ 1} + T_{CW\ 2}} \tag{2}$$

where $T_{Y-branch\ 1,2}$ are numerically computed power transmittances of the two stand-alone branches of a Y splitter (see Supplementary Note 3). The maximum power of THz light at 140 GHz operational frequency transfer from Waveguide 1 to Waveguide 2 is achieved by a 16.5 mm-long coupler, while the power is cycled back into Waveguide 1 in a 40.6 mm-long coupler. The corresponding electric field distributions of the two directional coupler circuits on the plane that is marked by violet dashed line in Fig. 3(a) are shown in Fig. 3(d).

Compared to the beat length between the two supermodes ($2L_c \sim 49$ mm) which is computed theoretically using 2D model solver, we find that the full-cycle coupling is realized by the directional coupler circuit containing a side coupler with a somewhat smaller length (40.6 mm) (see Fig. 3(e)). This difference mainly stems from the fact that coupling action between two waveguides persists over some distance in Y splitters as in this element the curved and straight waveguides run almost in parallel to each other over a certain distance. Finally, the incomplete power transfer in the coupler is due to significant loss difference of the two supermodes that have different field overlap with the lossy resin cage (see Supplementary Note 2). Additionally, due to asymmetric structure of Y splitters, the directional coupler circuit is no longer symmetric, thus breaking the phase matching condition between the two co-propagating modes that is required for the complete power transfer.

Experimental characterization of such circuits confirms sinusoidal behavior of the power coupling coefficients as a function of the side coupler length (see Fig. 3(e)). The experimentally found maximum power ratio for Waveguide 1 was obtained by a 35mm-long side coupler section, which is somewhat shorter than the predicted one. This is most probably related to the deviation of the experimental geometry of Y splitters from the ideal ones used in numerical modeling. As seen from the Insert of Fig. 3(b), due to limited resolution of 3D printer, the two waveguides remain joint for several mm at the entrance of a Y splitter, which results in stronger than expected inter-waveguide coupling in the fabricated Y splitters.

Due to the contribution of Y splitters to coupling between the bus and dropping waveguides, we design the ADM circuit using a 35mm-long side coupler section that enables the full-cycle coupling at ~140 GHz operational frequency with a broadband operation of over 60 GHz owing to the modest variation in the coupling length (see Supplementary Note 2).

## Grating-loaded side couplers for ADMs

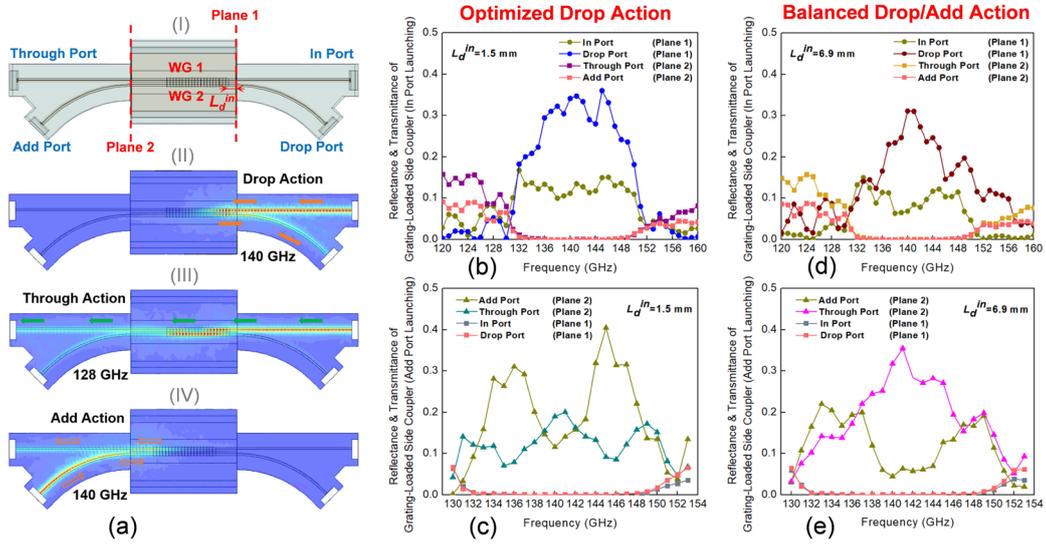

Fig. 4 Numerical study on the grating-loaded side coupler sections. (a) (I) Schematic of the ADM circuit used in numerical simulations of the experimental ADM shown in Fig. 1. It is composed of two Y splitters and a 35mm-long grating-loaded side coupler containing 20 periods with $\Lambda$=1.03 mm, $H$=0.21 mm. The computed electric field distributions in the ADM circuit with optimized drop action ($L_d^{in}$=1.5 mm) when using the fundamental mode of In Port (II) at 140 GHz (within grating stopband), and (III) at 128 GHz (within grating passband) as the excitation condition, and (IV) that in the ADM circuit with balanced drop/add action ($L_d^{in}$=6.9 mm) when using the fundamental mode of Add Port at 140 GHz (within grating stopband) as the excitation condition. Numerical transmittances and reflectance (by power) of grating-loaded side coupler with $L_d^{in}$ =1.5 mm at Planes 1 and 2 of the ADM with optimized drop action when use the fundamental mode of (b) In port (Plane 1), and (c) Add port (Plane 2) as the excitation condition. Numerical transmittances and reflectance (by power) of grating-loaded side coupler with $L_d^{in}$ =6.9 mm at Planes 1 and 2 of the ADM with balanced drop/add action when use the fundamental mode of (b) In port (Plane 1), and (c) Add Port (Plane 2) as the excitation condition. Figs. 4(b-e) are computed using Eq. (1) and numerical 3D models shown in Fig. 4(a).

The THz ADM circuits feature a grating-loaded side coupler placed between two Y splitters. Particularly, the directional coupler circuit enabling full-cycle coupling is modified by inscribing two Bragg gratings on a single set of two joint wires of side coupler section (see Figs. 1 and 4(a)). While propagating in the grating-loaded side coupler, the THz signal within the grating stopband that is launched into In port on

Waveguide 1 (Add port on Waveguide 2) is back-reflected by the gratings, and then transferred to Waveguide 2 (1) by the side coupler. The THz signal within the grating passband when launched into In port propagates through the grating-loaded side coupler and into Through port. Within the coupler bandwidth, the variation in the coupling length is moderate (see Fig. 3(c)), thus allowing high signal amplitude to be recorded at Through port. Separate numerical and experimental studies of Y splitters (see Supplementary Note 3) reveal that in such structures THz light mostly propagates in the forward direction regardless of the launching port with only a minimal amount of back-reflections and cross-talk (larger than 30 dB by power for any port). When integrating Y splitters into ADM circuits, their contribution to the crosstalk is negligible, with the main negative effect being additional propagation loss incurred by the signals propagating through the curved arms. Therefore, instead of using the actual powers at the different ports of ADM circuits after propagation through Y splitters, it is more convenient for the design purposes to use powers right at the output of grating-loaded side coupler element (Planes 1 and 2), which can be estimated by using Eq. (1) in numerical simulations (see examples in Fig. 4(a) and Supplementary Note 4 for detailed information).

It should be noted that the relative position of the 20.6mm-long gratings inside the 35mm-long side coupler section (controlled by the $L_d^{in}$ parameter) gives rise to different interference conditions between the two supermodes of a coupler, which we use to find optimal coupler designs. Particularly, to optimize the drop action we choose $L_d^{in}$ that results in the highest power at Drop port (Plane 1) within the grating stopband. Defining $L_g$ to be the light penetration distance into gratings, the desired relative position of the grating section $L_d^{in}$ can be found from:

$$L_d^{in} + L_g \approx L_c/2 \qquad (3)$$

Another optimal ADM design is to reduce the difference between spectral performance of the add and drop action in expense of their absolute performance, thus resulting in the balanced ADM. In this optimization, one minimizes the difference between transmission through the Through port under Add port launching and transmission through the Drop port under In port launching.

By comparing the ADM circuits featuring different $L_d^{in}$, the grating-loaded side couplers featuring $L_d^{in}$ of 1.5 mm and 6.9 mm are chosen to enable the ADM circuits with optimized drop action and balanced drop/add action, correspondingly. Figs. 4(b-e) show the numerically computed transmittance and reflectance of these two grating-loaded side couplers when the light is launched at In port (Plane 1) and Add port (Plane 2). We note that $L_g$ and $L_c$ parameters are frequency-dependent across the grating stopband. Therefore, ripples in transmittance and reflectance spectra of various ports are observed even for these optimal designs.

These two ADM designs were then experimentally characterized using CW-THz spectroscopy system. Firstly, we present the normalized power transmittances at Planes 1 and 2 to characterize the performance of the grating-loaded side coupler section while decoupling it from the losses incurred in Y splitters of an ADM. By using lower loss Y splitters one can significantly improve the loss characteristics of complete ADM devices, which are ultimately limited by those of the grating-loaded side coupler sections. While in this work, we use relatively lossy Y splitters comprising straight and curved branches. Further research is to develop more performant splitters, which is, however, beyond the scope of this paper.

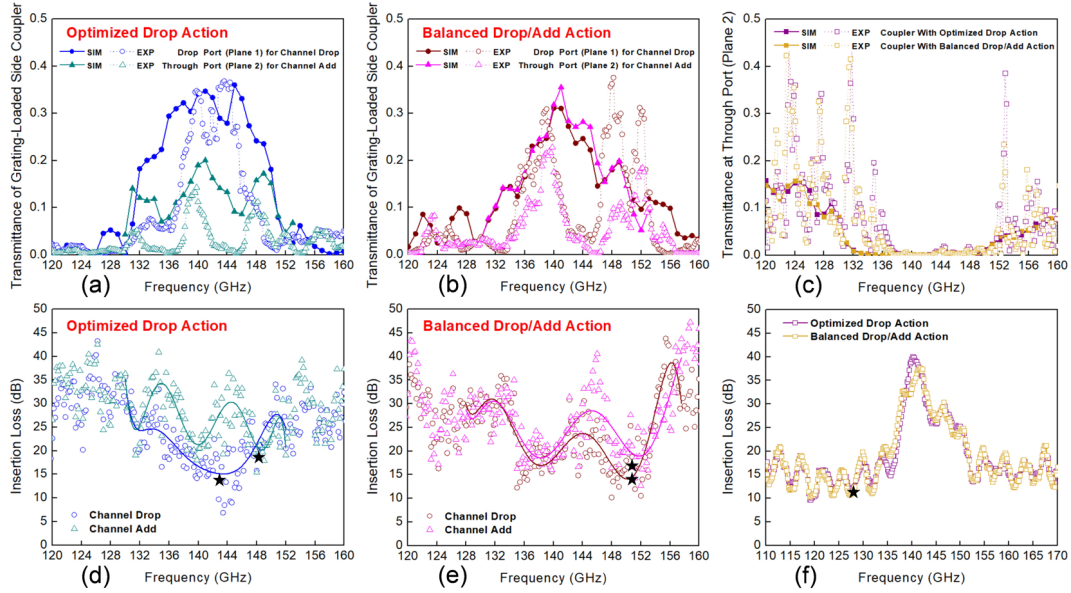

Fig. 5 Power transmittances of grating-loaded side couplers and insertion losses of complete ADM circuits. Numerical simulation and experimental results for the power transmittances of the grating-loaded side couplers that enable (a) the ADM with optimized drop action ($L_d^{in}$=1.5 mm) and (b) the ADM with balanced drop/add action ($L_d^{in}$=6.9 mm) under In port launching (Plane 1) and Add port launching (Plane 2) conditions. (c) Numerical simulation and experimental results for the power transmittances of these two grating-loaded side couplers at Through port (Plane 2) under In port launching (Plane 1) condition. The measured insertion losses (by power) under In (Add) port launching conditions as registered at Drop (Through) port of the complete ADM circuit (see Fig. 1) exhibiting (d) optimized drop action, and (e) balanced drop/add action. Solid lines are the 8th degree polynomial fits. (f) The measured insertion losses (by power) under In port launching condition as registered at Through port of the complete ADM circuits.

In Fig. 5(a) we present experimental and numerical transmittances at Drop port (Plane 1) under In port launching (Plane 1) and transmittances at Through port (Plane 2) under Add port launching (Plane 2) of the grating-loaded side coupler that enables the ADM circuit with optimized drop action. While an overall good correspondence between numerical and experimental data is observed, we also note that experimental transmittances show more pronounced ripples in their spectra, as well as somewhat lower bandwidths. This can be attributed to the suboptimal performance of the imperfect Bragg gratings due to manufacturing imperfections, as well as under-optimized (with respect to the Drop port) functioning of the Add port.

Additionally, compared with the numerical Bragg gratings featuring ideal geometry, the geometrical nonuniformity of the fabricated structure leads to the variation in the frequency-dependent $L_g$. As a result, it gives rise to the discrepancy in the interference of THz signal within the grating stopband propagating in the numerically and experimentally studied ADM circuits (Eq. (3)), thus resulting in the different transmittance of Drop port under In port launching (Through port under Add port launching). This inconsistency becomes pronounced for the grating-loaded side coupler designed for the ADM circuit with balanced drop/add action (see Fig. 5(b)). However, it is noted that similar transmittances were measured at Drop port (Plane 1) for channel dropping and Through port (Plane 2) for channel adding in experiments. Furthermore, Fig. 5(c) shows that the measured power transmittances at Through port (Plane 2) under In port launching (Plane 1) condition for these two grating-loaded side couplers in optimal designs are similar.

When inserting the complete ADM circuits containing these two grating-loaded side couplers (see Fig.

1) into two-wire waveguide-based THz communication networks, the insertion losses of THz light registered at different ports under In and Add port launching conditions are shown in Figs. 5(d-f) (see Supplementary Note 5 for details).

## Characterization of ADMs using THz communication system

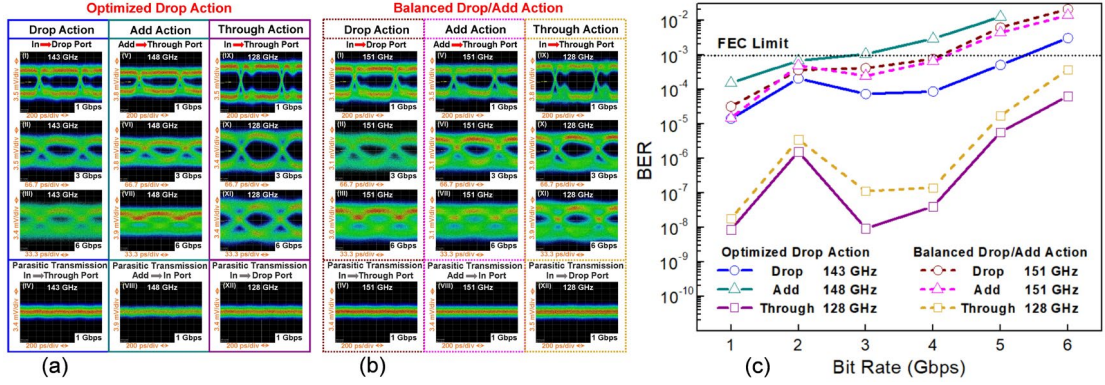

Fig. 6 Experimental study of two-wire ADM circuits for THz communications. (a) Eye patterns for the THz signals of different bit rates and different carrier frequencies propagating through the THz ADM circuit with optimized drop action. Measurements at (I-III) Drop port and (IV) Through port (In port launch, 143 GHz carrier), (V-VII) Through port and (VIII) In Port (Add port launch, 148 GHz carrier), as well as (IX-XI) Through port and (XII) Drop port (In port launch, 128 GHz carrier). (b) Eye patterns for the THz signals of different bit rates and different carrier frequencies propagating through the THz ADM circuit with balanced add/drop action. Measurements at (I-III) Drop port and (IV) Through port (In port launch, 151 GHz carrier), (V-VII) Through port and (VIII) In Port (Add port launch, 151 GHz carrier), as well as (IX-XI) Through port and (XII) Drop port (In port launch, 128 GHz carrier). The eye patterns featuring obvious eye opening were obtained by characterizing the response of these two ADMs at different ports in the spectral positions that are marked by black stars in Figs. 5(d-f). (c) Measured BER versus bit rate of the THz carrier wave propagating through two ADM circuits.

Finally, we experimentally characterized the two THz ADM circuits, one with optimized drop action ($L_d^{in}$ =1.5 mm) and the other one with balanced add/drop action ($L_d^{in}$ =6.9 mm), using an in-house photonics-based THz communication system with amplitude-shift-keying modulation format. The eye patterns are recorded at different ports of ADM circuits for THz signals within the grating stopband (add/drop action), as well as within the grating passband (through action) for various data rates as shown in Figs. 6(a) and 6(b). Due to relatively high insertion loss of our devices and low power of the optical THz source, the operational data rates are limited to ~6 Gbps.

For the ADM with optimized drop action, the minimal insertion loss (~14 dB) for channel dropping is observed in the middle of grating stopband at 143 GHz (see Fig. 5(d)). Therefore, when the THz signal with the carrier frequency of 143 GHz is launched into In port, the eye pattern recorded at Drop port shows the highest eye amplitude within the grating stopband (see Fig. 6(a) I-III), while the eye diagram for the parasitic transmission into Through port is shown in Fig. 6(a) IV. As only the drop action has been optimized, performance of the add action of this ADM is not symmetric with respect to that of the drop action. Thus, the highest efficiency of the add action is achieved within the grating stopband at 148 GHz (insertion loss of ~18 dB). In experiments THz signal was launched into Add port and measured at Through port for different data rates (see Fig. 6(a) V-VII), while parasitic transmission into In port is shown in Fig. 6(a) VIII. The corresponding BERs for the added (Through port) and dropped (Drop port) signals are shown in Fig. 6(c), from which it is clear that a nonsymmetric ADM also features nonsymmetric BER performances for the add and drop actions.

For the THz ADM with balanced drop/add performance, the lowest insertion loss (~15 dB) is at 151 GHz within the grating stopband (see Fig. 5(e)). Unsurprisingly, the resultant eye patterns for add and drop functions are similar and so are the corresponding BER curves (see Figs. 6(b) I-III & V-VII and 6(c)). In Fig. 6(b) IV and VIII we also show the parasitic transmission in Through port for the drop action and In port for the add action, correspondingly.

Despite the difference in the channel dropping and adding functions of these two ADMs, the through function of both ADMs in grating passband is remarkably similar (see Fig. 5(f)). This is manifested by the similar eye patterns (see Figs. 6(a) and 6(b) IX-XI) and BER curves (see Fig. 6(c)) for THz signals within the grating passband (ex. 128 GHz). Such similar performance in the passband is expected as the main difference between these two designs is the relative position of gratings within the side coupler sections of a fixed length. In the passband, though, transmitted light will only see the total length of the coupler circuit, while being somewhat modified by the presence of gratings. Parasitic transmissions into Drop port when the signal in grating passband is launched into In port for through action are shown in Figs. 6(a) and 6(b) XII. It is found that the crosstalk for the drop/add function in the stopband and through function in the passband is below the system noise level.

Finally, we present the BER response of these two ADM circuits (see Fig. 6(c)). In analysis we have to distinguish the through action in grating passband from the drop/add actions in grating stopband. For the through action, it is the dispersion and transmission losses of the straight waveguide sections along the In – Through path of ADMs, as well as dispersion and losses of the Bragg gratings in the passband that limit the data rates. For the drop/add action, it is the dispersion and losses in the straight and curved branches of a Y splitter, as well as dispersion and losses of the Bragg gratings in the stopband that limit the data rates. Generally, modal dispersion leads to signal scrabbling that manifests itself in the eye diagram distortion and closing of the eye pattern. From Fig. 6(c) we conclude that the through action in grating passband can support signal bandwidth greater than 6 Gbps as the maximal bitrate is mainly limited by the dispersion of a straight waveguide (assuming operation away from the edge of grating stopband). At the same time, it is also clear that the signal bandwidth for the drop/add action is rather limited by the size of grating stopband, which limits useful bitrates to ~2-5 Gbps (see Supplementary Note 6 for details).

Additionally, losses of various ADM subcomponents significantly affect the BER of the device due to negative effect of the noise on the detection accuracy. We observe that the highest loss in our device ~9 dB comes from the 6.3cm-long curved branch of a Y splitter. This loss is mainly due to radiation from a curved waveguide and ohmic losses of plasmonic modes. Next, the grating-loaded side coupler section typically results in ~7 dB loss for operation in grating passband (for through action), as well as >4 dB loss for operation in grating stopband (for drop and add action). This loss is mainly due to mismatch between the modal fields of straight two-wire waveguides and gratings, ohmic losses, as well as imperfections in the grating structure. Next, there is butt-coupling loss of ~3 dB at each interface between ADM and waveguide-coupled THz emitter and detector [28], which has not been counted in Figs. 5(d-f). Finally, there are relatively low losses of straight waveguides (0.4 dB/cm) which are mostly due to ohmic losses in the deposited silver layer incurred by the plasmonic modes. As a result, our ADMs feature relatively high insertion losses and modest signal-to-noise ratio (see Supplementary Note 7 for details).

# 3. Discussion

We demonstrated four-port ADM circuits capable of channel adding and dropping for applications in THz communication systems operating within FDM modality. We used 3D-printed two-wire plasmonic waveguide as a base technology to develop reconfigurable plasmonic circuits comprising WBGs, directional couplers, grating-loaded side couplers, and Y splitters. We believe that the air-core two-wire waveguides offer many advantages for THz circuit development, including: low losses due to high modal presence in air, device tunability due to ease of access to modal fields, 3D integration potential, modular nature and low coupling loss between modules, as well as capability for robust and cost-effective prototyping and fabrication. Two ADM types were studied, one is a drop action optimized design, the other one is a balanced drop/add design, both featuring experimental bandwidths of 12 GHz (grating stopband) in the vicinity of 143 GHz. For the drop optimized ADM circuit, drop action with data rates of up to 5 Gbps at 143 GHz carrier and add action with data rates of up to 3 Gbps at 148 GHz carrier were confirmed to result in bit error rates lower than $10^{-3}$ (forward error correction (FEC) limit). For the balanced add/drop multiplexer design, both add and drop actions supported data rates of up to 4 Gbps at 151 GHz carrier with BER<$10^{-3}$. Through action in the grating passband (ex. 128 GHz) was shown to support bitrates exceeding 6 Gbps for both designs.

The main challenge for the proposed ADM designs was relatively high insertion losses (~10 dB for through action and larger than 14 dB for drop and add action) together with the butt-coupling loss of ~6 dB. Main reasons for such losses, in order of importance, are: suboptimal Y splitter (~9 dB), insertion losses of Bragg gratings and scattering in gratings due to fabrication imperfections (~7 dB for grating passband and >4 dB for grating stopband), butt-coupling loss at each interface (~3 dB), transmission losses of straight two-wire waveguides (~0.4 dB/cm). We believe that further work on relatively straightforward optimization of ADM circuits, including employing more efficient splitter design and better fabrication strategies, as well as reducing coupling losses to detector and emitter by geometry optimization of output ports, could reduce the insertion loss by 5-10 dB. It can also improve and balance the performance of add and drop actions, thus rendering ADM devices highly suitable for the ultra-fast analogue signal processing in THz communications.

# Methods

**Fabrication of two-wire waveguide components**

For the ease of fabrication, each proposed micro-encapsulated two-wire plasmonic waveguide component such as straight and curved waveguide section, WBG module, directional coupler, and grating-loaded side coupler, is split into two complementary parts. Each part comprises of one or two joint wires (depends on the waveguide component) that are attached to a half cage using deeply subwavelength dielectric support ridges of width 150 μm (see Figs. 1, 2(a), and 3(a)) [28]. These structures were then fabricated using SLA 3D printer (Asiga® Freeform PRO2) that features XY-axis resolution of 50 μm. It is noted that the length of the waveguide is fabricated along the Z-axis of the 3D printer. To ensure high geometrical precision, the parts of the waveguide component with periodic variation in cross sections were printed using the finest resolution in the Z-axis (layer thickness of 10 μm). While the parts with smooth surfaces were fabricated with the Z-axis resolution of 25 μm to reduce the fabrication duration and moreover it does not affect the overall performance of the component. After fabrication, the inner surface of the cage was covered using masking tape leaving the plastic wire uncovered. Then, silver layer was deposited on top of the plastic wires using wet chemistry deposition to form conductive surfaces. Finally, the two-wire waveguide components were obtained by assembling the two selectively metallized parts by sliding the corresponding parts into each other using V-grooves and V-ridges that were printed onto the cage. The proposed THz circuits (directional coupler and ADM circuits) were assembled from these two-wire waveguide components using another set of alignment and connectorization elements that are imprinted onto the cage end facets for seamless connectorization (see Figs. 1 and 3(b)).

**Characterization of two-wire WBGs**

The proposed micro-encapsulated two-wire WBGs were characterized using the CW-THz spectroscopy system (Toptica Photonics) [48-50]. The schematic of the experimental setup is shown in Supplementary Note 8 and briefly explained as follows. Two distributed feedback (DFB) lasers with power of ~30 mW each, operating in the infrared C-band with slightly different emission wavelengths are used as the source of THz generation. A 50:50 fiber coupler is used to combine the two laser beams and split into both emission and detection arms respectively. By applying AC bias voltage to the emitter photomixer, the THz radiation corresponding to the beat frequency between the two DFB lasers is generated. The output THz frequency can be varied by simply tuning the emission wavelengths of the lasers. A similar photomixer (without bias voltage) and a lock-in amplifier are used in THz detection arm. The focused linearly-polarized CW-THz beam transmitting through the waveguide components is re-collimated and the amplitude of the THz signal is recorded using lock-in detection. The phase of the THz signal is simultaneously recorded by using the fiber stretchers of equal lengths that are connected to both emitter and detector arms.

The relative transmission coefficient of WBG was characterized as follows. Firstly, a two-wire waveguide section featuring same length with the WBGs (2.5 cm) was placed between two waveguide sections (3 cm). The assembled two-wire waveguide was connected with WR6.5 conical horn antennas (Virginia Diode, Inc.) at its both ends, and then inserted into the THz beam path of the spectroscopy system with the input and output planes of the waveguide flanges at the focal points of plano convex lenses (PCL1 and PCL2). It is noted that, a metal barrier is placed around the horn antenna at the input port to block the residual THz signal entering the detector. Then, the transmission spectrum (by field) of the assembled waveguide was measured as the reference. Next, the 2.5cm-long waveguide section was replaced by the WBG element, and the transmission spectrum of the assembled waveguide component was recorded. The relative transmission coefficient (by field) of WBG was finally obtained by dividing the measured spectrum with waveguide sections by the reference.

The relative reflection coefficient of WBG was characterized as follows. Firstly, the two-wire Y-coupler and the WBG were assembled using the interconnects at the end facets of both components. Then, two WR6.5 conical horn antennas were connected to the unused ports, namely input and output ports of the Y-coupler. The THz light was launched into the input port of Y-coupler, guided through a curved arm of the coupler, and reflected from the WBG under study. The reflected light was then divided equally by the Y junction and directed towards both the input and output ports. The transmission spectrum of THz light at the output port was recorded. Next, the WBG was replaced by a plano metallic mirror to completely reflect the THz signal at the Y-junction. The transmission spectrum recorded at the output port was used as the reference. Finally, the relative reflection coefficient (by field) of WBG was obtained by comparing these two measured transmission spectra.

## Characterization of side coupler section

The THz side coupler containing two two-wire waveguides was characterized as follows. Firstly, the side coupler was placed between two Y splitters (see Fig. 3(b)). Each port of the assembled THz circuit was connected to a WR6.5 conical horn antenna, and then inserted into the CW-THz spectroscopy system (see Supplementary Note 9). Initially, the THz signal was launched into IP 1 and the transmitted power received at OP 2 was recorded. Then, by rotating the rotary section by 90° in the clockwise direction, the transmitted power at OP 1 was recorded. The power of THz light in the Waveguide 2 and 1 at output plane of the side coupler section (Plane 2) (see Fig. 3(b)) was obtained by dividing the transmitted power recorded at OP 2 and OP 1 by the insertion loss of the curved and straight two-wire waveguide in the two arms of Y splitter (see Supplementary Note 3), correspondingly. The power ratio of each waveguide was then computed by dividing its derived output power by the sum of these two values. By subsequently abrading the side coupler section to reduce its length and repeating the similar measurements, the power ratios of the two two-wire waveguides of straight side coupler sections featuring different lengths were obtained.

## Characterization of two-wire ADM circuits using CW-THz spectroscopy system

The transmittance (by power) of grating-loaded side couplers and the insertion loss (by power) of complete THz ADM circuits (see Fig. 1) were characterized using CW-THz spectroscopy system. The grating-loaded side coupler section was placed between two Y splitters. Each port of the assembled THz ADM circuit was then connected with a WR6.5 conical horn antenna. Firstly, to measure the transmittance of Drop port under In port launching at Plane 1 of grating-loaded side coupler for channel dropping, the ADM circuit was inserted into the CW-THz spectroscopy system with the input and output planes of waveguide flanges connected with In and Drop ports at the focal points of PCL1 and PCL2 (see Supplementary Note 9). The transmitted power spectrum of Drop port was recorded when the THz light was launched into In port. Next, the ADM circuit was replaced by a two-wire waveguide assembled by straight and curved waveguide sections which features the same length with the sum of the two arms of a Y splitter. Its transmitted power spectrum was recorded as the reference. The transmittance of Drop port at Plane 1 of grating-loaded side coupler was then obtained by dividing the measured transmitted power spectrum at Drop port of the complete THz ADM circuit by this reference spectrum to decouple it from the losses incurred in the Y splitter.
Next, the reference spectrum that is used to characterize the insertion loss of a complete THz ADM circuit when it is inserted into two-wire waveguide-based communication system was obtained by dividing the measured transmitted power spectrum of a straight two-wire waveguide by its transmission loss [28]. The insertion loss (by power) of the THz carrier wave for channel dropping was then computed by dividing the measured transmitted power spectrum at Drop port of THz ADM under In port launching condition by this reference. The transmittance of Through port of grating-loaded side coupler under Add port launching condition at Plane 2 for channel adding and the corresponding insertion loss of complete ADM circuit were characterized using the similar processes after placing the input and output planes of waveguide flanges connected with Add and Through ports at the focal points of PCL1 and PCL2 in CW-THz spectroscopy system.

To characterize the transmittance of Through port at Plane 2 of the grating-loaded side coupler section under In port launching (Plane 1) condition, the THz ADM circuit was placed inside the CW-THz spectroscopy system with the input and output planes of waveguide flanges connected with In and Through ports at the focal points of PCL1 and PCL2. Then the transmitted power spectrum of Through port was recorded when the THz light was launched into In port. Next, the transmitted power spectrum of a two-wire waveguide having a same length with the sum of straight arms in two Y splitters was recorded as reference spectrum. The transmittance of Through port at Plane 2 was obtained by dividing the transmitted power spectrum recorded at Through port of ADM by this reference. The insertion loss of the THz carrier wave propagating from In port to Through port of the complete ADM circuit was then computed by comparing the transmitted power spectrum recorded at Through port and the reference spectrum which has been used to measure the insertion loss of ADM for channel dropping and adding.

## Characterization of ADM circuits using THz communication system

The two-wire plasmonic ADM circuits were also characterized using an in-house photonics-based THz communication system. The schematic diagram of the experimental setup is shown in Supplementary Note 10 and briefly explained as follows. In the transmitter section, two independently tunable DFB lasers (Toptica Photonics) operating in the infrared C-band with slightly different center frequencies are combined using a 3 dB coupler as the source of THz generation. The baseband signal source of pseudorandom bit sequence (PRBS) with a varying bit rate from 1 Gbps to 6 Gbps and pattern length of $2^{31} - 1$ is generated by the pulse pattern generator (PPG) unit integrated in the test equipment (Anritsu-MP2100B). The baseband signal is then amplified using the RF amplifier (Thorlabs-MX10A) which drives the Mach–Zehnder modulator (Thorlabs-LN81S-FC) to modulate the intensity of laser beams. Then the modulated laser beams are amplified using an erbium-doped fiber amplifier (EDFA) (Calmar laser-AMP-PM-18) and injected into a waveguide coupled uni-traveling-carrier-photodiode (UTC-PD) photomixer (NTT Electronics) to generate modulated THz carrier wave with operation frequency corresponding to the beat frequency of the two tunable DFB lasers. In the receiver section, the THz carrier wave is detected and demodulated by a zero bias Schottky diode (Virginia Diodes-WR6.5-ZBD-F), and then amplified using a high gain low noise amplifier (LNA) (Fairview Microwave-SLNA-030-32-30-SMA). Finally, the eye pattern and BER are recorded using the test equipment (Anritsu-MP2100B) [51,52].

In experiments, the THz carrier wave within the spectral range of 110 -170 GHz having the power in the range between 125 μW and 250 μW was butt-coupled into the THz ADM circuit via a 1-inch-long WR6.5 rectangular waveguide (Virginia Diodes WR6.5). A similar arrangement was used to connect the output port of the THz ADM circuit and the Schottky diode (see Supplementary Note 10). We switched the ports of ADM circuits connected with the THz transmitter and receiver, and characterized the THz carrier wave featuring operation frequencies in the stopband and passband of gratings that are loaded on side coupler section subsequently. Three scenarios, that are, the transmitter and receiver are connected to In and Drop ports, In and Through ports, as well as Add and Through ports, correspondingly, were mainly studied. In each case, the eye pattern and the corresponding BER measurements of different bit rates were carried out for the carrier frequencies in the grating stopband and passband of the ADM circuits, respectively. It is noted that, during BER measurements, the decision threshold was optimized to equalize the insertion error (digital 0 is mistaken as digital 1) and omission error (digital 1 is mistaken as digital 0). It is then recorded within the duration of 1/(target BER× bit rate), where the target BER was set as 10$^{-12}$ (error-free).

**Numerical simulation**

The numerical studies of the proposed WBGs, directional coupler circuits, and ADM circuits were carried out using the commercial finite element software COMSOL Multiphysics within the 3D finite element frequency domain module with ports. Within this formulation, transmission and reflection can be characterized by using scattering matrix coefficients (S-parameters) related to each defined port. In our simulations we used frequency-dependent refractive index ($n_{resin}(f)$=1.654-0.07$f$ [THz]) and material absorption ($\alpha(f)$ [cm$^{-1}$]=0.64+13.44($f$ [THz])$^2$) of resin (used in 3D printing) in THz spectral range of 0.1-0.3 THz, which were obtained in a prior experimental study detailed in Ref. [53]. The metallic wires were modeled using impedance boundary condition (IBC) at the wire surface together with the Drude-Lorentz model for the dielectric constant of silver obtained from [28,54].

$$\varepsilon_m = 1 - \frac{\omega_p^2}{\omega^2 + i\omega\gamma_b} \tag{4}$$

where $\omega_p$=2π·2.185e15 Hz is the plasma frequency of silver, and $\gamma_b$=2π·2.69e14 Hz is the fitted value of the damping coefficient of the Ag layer deposited on resin support in THz spectral range.

Due to geometrical symmetry of the two-wire WBG module, only half of this structure together with perfect electrical conductor (PEC) boundary condition was used in numerical simulations. When the THz light is launched using port boundary condition at the input facet of the WBG, the reflection coefficient at the same port ($S_{11}$) and transmission coefficient at the port on the other end facet ($S_{21}$) were computed (see Supplementary Note 11 for example)

By following the similar procedure discussed above, the directional coupler and ADM circuits were studied using the numerical simulations. Different from the WBG models, the complete structure of the directional coupler and ADM circuits were studied using 3D models due to their asymmetric structures. The scattering matrix coefficient ($S^2$) at the four defined ports are computed under various launching conditions, i.e. OP 1 and OP 2 under IP 1 launching for directional coupler circuit (see Fig. 3(d)), as well as In, Drop, Add, and Through ports under In port launching and Add port launching conditions for ADM circuit (see Fig. 4(a)). The transmittance and reflectance (by power) of side couplers at Planes 1 and 2 were then obtained using Eq. (1).

# Data availability

The authors declare that the main data supporting the findings of this study are available within the article and its Supplementary Information files. Extra data are available from the corresponding author upon request.

# Funding

This work was supported by Pr. Skorobogatiy Canada Research Chair I in Ubiquitous THz photonics (34633); China Scholarship Council (201706250016).

# Contributions

All authors conceived these experiments and contributed to their design. Y. C and K. N performed the numerical simulations and measurements. Y. C, K. N, and M. S contributed to writing the manuscript.

# Corresponding author

Correspondence to Maksim Skorobogatiy.

# Ethics declarations

## Competing interests

The authors declare no competing interests.

# Add drop multiplexers for terahertz communications using two-wire waveguide based plasmonic circuits

Yang Cao, Kathirvel Nallappan, Guofu Xu, and Maksim Skorobogatiy[*]

*Department of Engineering Physics, École Polytechnique de Montréal, Montreal, Québec, H3T 1J4, Canada*
*[*maksim.skorobogatiy@polymtl.ca](mailto:maksim.skorobogatiy@polymtl.ca)*

# Supplementary Note 1. Design of two-wire waveguide Bragg gratings (WBGs)

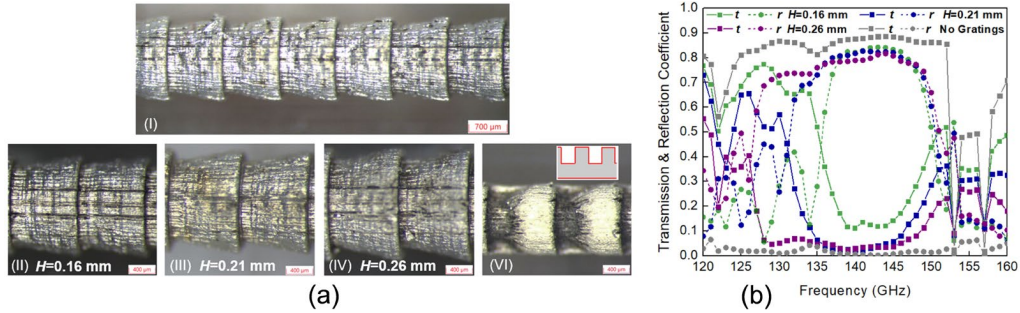

Fig. S1 The design of two-wire WBGs. (a) (I) Photographs from the top of a fabricated WBG with a period of $\Lambda$=1.03 mm and a ridge height of $H$=0.21 mm (only the WBG half is shown). (II-IV) Enlarged view of the metalized gratings (truncated cones) of different ridge heights $H$. (V) Photograph of a metalized grating featuring rectangular grooves and its schematic. (c) The numerical transmission and reflection coefficients (by field) of the 2.5cm-long WBGs containing 20 periods with $\Lambda$=1.03 mm as a function of the ridge height $H$.

The two-wire WBG features periodic variation in cross sections of one wire, while the other wire is uniform and has a fixed diameter of $D$=1.321 mm. The periodic structure is a sequence of end-to-end connected truncated cones added on top of an uniform circular wire of diameter $D$ (see Fig. S1(a) (I-IV)). The choice of truncated cones in the grating structure was experimentally found to be the most reliable and stable for printing compared to other alternatives such as rectangular or sinusoidal ridges, or rectangular grooves on wire surface. The deformations in these 3D printed structure is mainly attributed to the "cure-through" effect during 3D printing. Particularly, the UV radiation not only cures the resin within the top printed layer, but also leaks through the cured layer and solidifies some resin on the other side. This uncontrollable effect occurs repeatedly, thus resulting in the cumulative deformation of prints, which is most pronounced when geometry changes rapidly from one layer to another. For example, the ideal steep slope at the edge of a rectangular ~600um-deep groove engraved on a wire was deformed into a conical structure when printed (see Fig. S1(a) V). This problem highlights the importance of carefully choosing the photosensitive resin chemistry to match the UV absorption depth with the desired layer thickness. Finally, we mention that gratings used in this work can only be printed along a single direction from the conus smaller base towards its larger base to avoid overhanging structures.

In Fig. S1(b) we show computed transmission and reflection coefficients (by field) of THz light propagating through 2.5cm-long WBG sections as a function of the ridge height $H$, which effectively controls grating strength (stopband bandwidth). The gratings contain 20 periods of truncated cones with a grating period of $\Lambda$=1.03 mm. As a reference, transmission and reflection coefficients for a straight two-wire waveguide of the same length are also shown. Due to anticrossing of the waveguide fundamental modes with those of a resin cage, sharp dips in the reference waveguide transmission spectrum are present at 122 GHz and 155 GHz, while in-between these values waveguide transmission is relatively featureless. Similar anticrossing phenomenon is also observed numerically in the transmission spectra of WBGs. However, such spectral features are very sensitive to the surface roughness and other structural imperfections, and in fact not easily observed experimentally as reported earlier in Ref. [1]. We also note that increasing the ridge height $H$ from 0.16 mm to 0.26 mm results in strong increase in the size of grating stopband from ~12 GHz to ~25 GHz calculated as a full width at half maximum (FWHM) of the reflection coefficient curve. Additionally, increase in the ridge height results in a smaller average gap size between the two wires, and as a consequence, higher effective refractive index of a WBG mode, thus reducing the center frequency of the grating stopband. Considering the finite transverse resolution of the stereolithography (SLA) 3D printer (50 μm in our case), the truncated cone-shaped grating section featuring larger ridge heights (Fig. S1(a) IV) can be fabricated easier and more reliable than those with smaller heights (Fig. S1(a) II). However, at the same time, larger ridge heights require tighter tolerances for the various sizes of the supporting cage due to smaller separation between the grating wire and the uniform wire. Using larger ridge heights also tends to result in lossier gratings due to stronger scattering effect of imperfections in the grating structure on the guided modal fields. Therefore, an optimal ridge height of the gratings must be chosen to ensure large operational bandwidth, reproducible optical performance, and manageable loss. In this work, by comparing optical properties of several WBGs we find the optimal ridge height to be $H$=0.21 mm (see Fig. S1(a) I and III). With the period of a WBG set to $\Lambda$=1.03 mm, the resultant stopband center frequency is 140 GHz with the corresponding bandwidth of ~18 GHz.

**Supplementary Note 2. Numerical simulation results of directional coupler circuit**

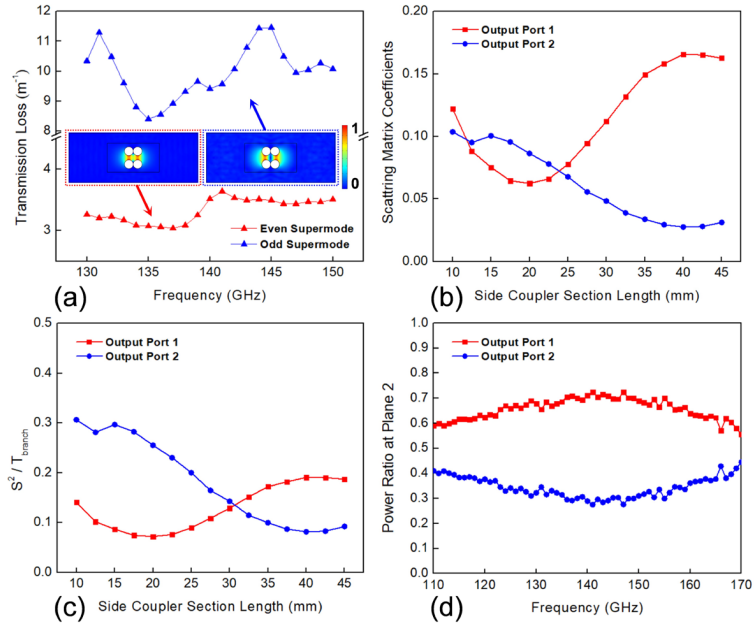

Fig. S2 Numerical simulation results of directional coupler circuit (a) The transmission loss (by field) of the even and odd supermode supported by the side coupler section that contains two two-wire waveguides. Insert: the modal electric field distribution of the two supermodes at 140 GHz operational frequency. (b) The numerically computed power transmittance at the Output port 1 and Output port 2 assuming excitation at the Input port 1 of the directional coupler circuit (see Fig. 3(d)) as a function of the length of the straight side coupler section. (c) The numerically computed power transmittance of the two waveguides (Waveguides 1 and 2) at the end (Plane 2) of side coupler section assuming excitation at Input port 1 of the directional coupler circuit (see Fig. 3(d)) as a function of the length of the straight side coupler section (obtained by using Eq. (1)). (d) The ratio of the power carried by the two two-wire waveguides (Waveguides 1 and 2) at the end (Plane 2) of a 40.6mm-long side coupler section when it is integrated into the directional coupler circuit shown in Fig. 3(d), whose value is computed using Eqs. (1) and (2).

**Supplementary Note 3. Characterization of Y splitters used in the directional coupler and ADM circuits**

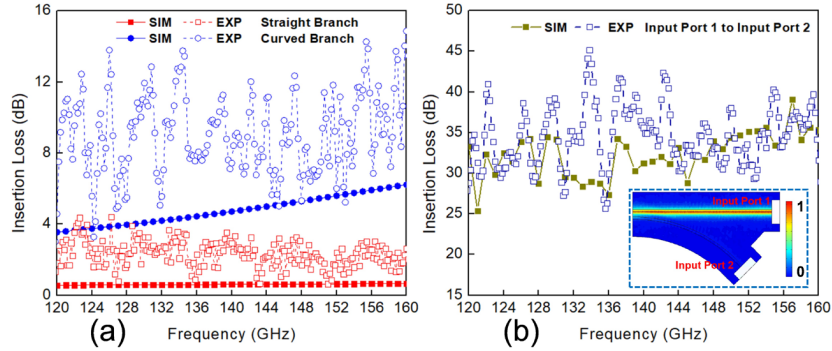

Fig. S3 Characterization of two-wire Y splitters. (a) The insertion loss of the straight and curved arms of the Y splitters used in the directional coupler and ADM circuits for numerical and experimental studies. (b) The insertion loss of the THz light that is launched into Input port 1 and registered at Input port 2 of Y splitters. Insert: the computed electric field distributions in the Y splitter when using the fundamental mode of Input port 1 at 140 GHz as the excitation condition in numerical simulations.

In this work, the two-wire waveguide-based broadband Y splitter (see Figs.1, 3(b), 3(d), and 4(a)) was used to guide THz signals to/from the side coupler section from/to the desired ports of directional coupler and ADM circuits. It features a curved two-wire waveguide (circular arc of radius 4 cm) gradually approaching a 4.5cm-long straight two-wire waveguide, while joining in a wedge configuration at the end. In this work, the side coupler sections were connected with Y splitters at their two ends to form the complete directional coupler and ADM circuits. In numerical simulation, the transmittance and reflectance of side coupler section were computed by dividing the scattering matrix coefficient ($S^2$) of the desired port of the complete circuit by the insertion loss of the arms in Y splitters. In experiments, the transmittance of the side coupler section was measured by dividing the transmitted power spectrum recorded at the desired port of the complete circuit by the transmitted power spectra of a two-wire waveguide having a length same with the arms in Y splitters which the THz carrier wave propagates through.

Additionally, Fig. S3(b) shows that in Y splitter, the THz light that is launched into one port mostly propagates in the forward direction (larger than 30 dB by power) with only a minimal amount of crosstalk to the other port. Therefore, when using Y splitters in directional coupler and ADM circuits, their contribution to the crosstalk is negligible.

## Supplementary Note 4. Numerical simulation of two-wire ADM circuits

In what follows we detail design and optimization of the ADM circuit shown in Figs. 1 and 4(a). The orange arrows represent THz signals that are within the stopband of a grating, while green arrows correspond to those in the grating passband. The path of the THz light that is launched into In port (Waveguide 1) is marked by solid arrows, while that launched into Add port (Waveguide 2) is marked by hollow arrows. While propagating in the grating-loaded side coupler, the THz signal within the grating stopband that is launched into In port on Waveguide 1 (Add port on Waveguide 2) is back-reflected by the gratings, and then transferred to Waveguide 2 (Waveguide 1) by the coupler. The THz signal within the grating passband when launched into In port propagates through the grating-loaded side coupler and into Through port.

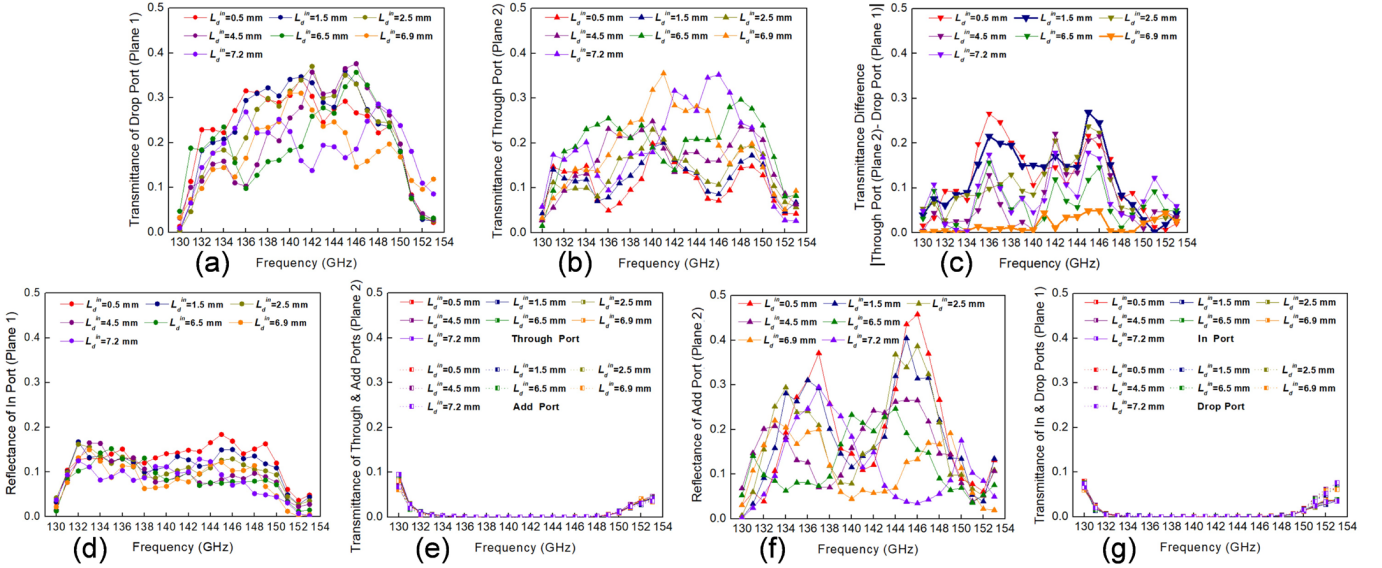

Fig. S4 Numerical simulation results of 35mm-long grating-loaded side coupler sections containing 20 periods with $\Lambda$=1.03 mm, $H$=0.21 mm, and different $L_d^{in}$ computed using Eq. (1). Numerical transmittances (by power) at (a) Plane 1 for Drop port under In port launching, and (b) Plane 2 for Through port under Add port launching conditions. (c) Minimizing the difference between transmittance of the Drop port at Plane 1 (drop function) and that of the Through port at Plane 2 (add function) to equalize the drop and add functions. Numerical reflectance (by power) at (d) Plane 1 for In port, and (f) Plane 2 for Add port of ADM circuits. Parasitic transmittance (by power) at (e) Plane 2 for Through and Add ports under In port launching (Plane 1), and (g) Plane 1 for In and Drop ports under Add port launching (Plane 2) conditions.

Firstly, we numerically study transmittance and reflectance (by power) characteristics of the 35mm-long grating-loaded side couplers containing 20.6mm-long gratings using Eq. (1) and scattering matrix coefficients ($S^2$) at the In, Drop, Add, and Through ports for various launching conditions. By varying the $L_d^{in}$ parameter, particular attention was paid to the design of two optimal grating-loaded side couplers that enable the ADM circuits with optimized drop and balanced drop/add actions, respectively.

Numerical simulations show that the maximum value of the transmittance of Drop port at Plane 1 for the operation frequency within the grating stopband ~ 140 GHz ± 6 GHz, which gives rise to the optimized drop action of ADM circuit, is obtained when $L_d^{in}$~0.5-2.5 mm (see Fig. S4(a)). This shows the realistic tolerances when choosing the value of $L_d^{in}$ for the optimized drop action considering that the SLA printer resolution is ~50 um.

For the design of the other optimal ADM with balanced drop/add action, one needs to minimize the difference between transmission through the Through port under Add port launching condition and transmission through the Drop port under In port launching condition. Particularly, we vary the $L_d^{in}$ parameter to minimize the absolute difference between transmittances of the Through and Drop ports (see Figs. S4(a-c)). Clearly, by choosing identical left and right grating offsets $L_d^{in} = L_d^{out}$ should result in the balanced performance of channel dropping and adding in case of the ADM with symmetric gratings [2,3]. In our case, we observe that the balanced ADM performance is achieved when $L_d^{out} - L_d^{in} = 0.6$ mm ($L_d^{in}$=6.9 mm). This somewhat off-center positioning of gratings is attributed to the small asymmetry in the grating performance depending on the direction of propagation of light.

Additionally, we also present analysis of parasitic transmissions into Through and Add ports (Plane 2) when light is launched into In port (Plane 1) within the grating stopband. In this case, most of the energy (~15-40%) is transferred into Drop port (Plane 1) (Fig. S4(a)), some is reflected back into In port (Plane 1) (~5-15%) (Fig. S4(d)), while parasitic transmissions to Through and Add ports (Plane 2) constitute <1% (Fig. S4(e)). Similar conclusions are reached for the four ports when light is launched into Add port (Plane 2) within the grating stopband (see Figs. S4(b), S4(f), S4(g)).

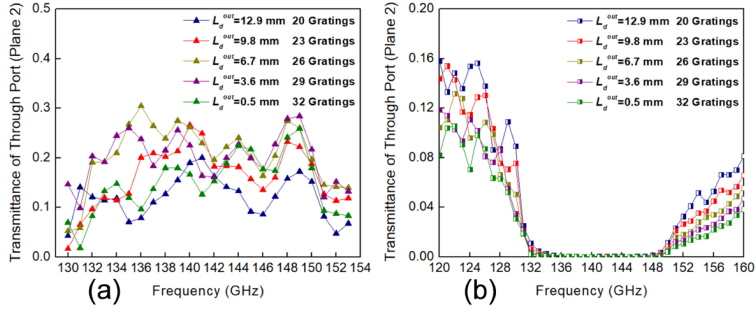

Fig. S5 Numerical transmittances (by power) of the 35mm-long grating-loaded side coupler sections containing gratings featuring different numbers of periods with $\Lambda$ =1.03 mm, $H$=0.21 mm, and $L_d^{in}$ =1.5 mm computed using Eq. (1) at Plane 2 for Through port under (a) Add port launching (Plane 2), and (b) In port launching (Plane 1) conditions.

For the above-mentioned grating-loaded side couplers, the value of $L_d^{out}$ might be suboptimal as it is not truly a free parameter. In fact, if $N_g$ is the number of grating periods, then:

$$L_d^{out} = L_{coupler} - L_d^{in} - N_g \Lambda \qquad (S1)$$

Since $L_d^{in}$ has been determined as 1.5 mm to enable the optimized drop action, $L_d^{out}$ can be adjusted by varying the number of grating periods to optimize add action simultaneously. One usually can set $L_d^{out} \approx L_d^{in}$, which requires the use of longer gratings (ex. $N_g$=30) (see Fig. S5). However, we found experimentally that longer Bragg gratings $N_g$ ~22-31 result in high transmission losses due to scattering loss on imperfectly printed grating structure. At the same time, one requires at least $N_g$ =15-20 in experiments to make sure that the grating length is longer than the light penetration depth into the grating $L_g$. Thus, for the ADM design with imperfect gratings there is a tradeoff between the equivalently optimized performance of channel dropping and adding and the ADM loss. In experiments, we prioritized low ADM losses while somewhat sacrificing the optimized operation of both of Add and Drop ports by choosing $N_g$ =20, which is mostly manifested by the elevated back reflection in the case of Add port (Plane 2) (see Fig. S4(f)) which is more pronounced than back reflection into In port (Plane 1) for the drop action (see Fig. S4(d)).

**Supplementary Note 5. Measured insertion loss of two-wire ADM circuits**

When inserting the complete ADM circuit (see Fig. 1) into two-wire waveguide-based THz communication networks, the insertion losses of THz light registered at different ports under In and Add port launching conditions are shown in Fig. 5(d-f). For the ADM with optimized drop action ($L_d^{in}$ =1.5 mm), the minimum insertion loss by power (within the Bragg grating stopband) as registered at Drop port under In port launching was measured as 14 dB at 143 GHz. While for the Through port under Add port launching, the minimal insertion loss was 18 dB at 148 GHz for channel adding (see Fig. 5(d)). Similarly, for the ADM circuit with balanced drop/add performance ($L_d^{in}$ =6.9 mm), the minimal insertion losses were measured at two spectral locations ~138 GHz and ~151 GHz with the corresponding values of ~16 dB and ~15 dB for Drop port under In port launching and ~18 dB and ~15 dB for Through port under Add port launching. Besides, an insertion loss peak appears around the center of grating stopband on the fitted spectra (see Fig. 5(e)). Finally, insertion losses for the two ADM designs as registered at Through port under In port launching are shown in Fig. 5(f). The pronounced loss peaks corresponding to the drop action of ~12 GHz FWHM bandwidth are clearly observable in the vicinity of ~140 GHz.

## Supplementary Note 6. Influence of dispersion of ADM on data transmission

To quantify the effect of the link dispersion on the maximal supported bit rate, we consider second order modal dispersion $\beta_2$. The maximum bit rate '$B$' (for ASK modulation) supported by the waveguide of a length '$L$' can be estimated using Eq. (S2), which is derived by requiring that ~95% of the power of the broadened pulse form still remains within the time slot allocated to logical "1" [4]:

$$B = \frac{1}{4\sqrt{|\beta_2|L}} \tag{S2}$$

In [1] it was demonstrated that in the 120-160 GHz spectral range, the two-wire waveguides (same as used in the ADMs of this paper) feature low GVD which are < 3 ps/(THz*cm) for the straight waveguide and <10 ps/(THz*cm) for the bend waveguide with 4 cm bending radius. Thus, according to Eq. (S2), dispersion of the 12.5cm-long straight In-Through waveguide will only affect data rates exceeding 40 Gbps. It is noted that in this estimate we assumed that dispersion of a grating in the passband is similar to that of a straight waveguide, which is a valid assumption for the operational frequency far away from the edges of grating stopband.

In contrast, for the drop/add action in the grating stopband, reflective properties of the Bragg grating vary greatly within the stopband. Ultimately, it is the spectral size of the grating stopband that limits the signal bandwidth. Furthermore, due to fracturing of the stopband because of the fabrication imperfections and various interference effects (see Figs. 5(d) and 5(e)), signal bandwidth for the drop/add action in our ADMs is expected to be limited to ~2-4 GHz.

**Supplementary Note 7. Signal-to-noise ratio (SNR) of THz ADM circuits**

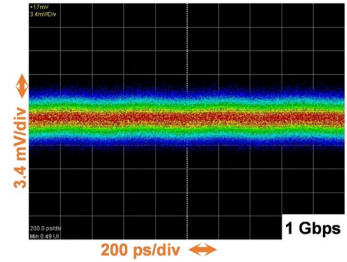

Fig. S6 Eye diagram for the noise level of the THz communication system

Fig. S6 shows the eye diagram for the noise level (no signal) of the THz communication system. Due to the relatively high insertion losses for both the drop/add actions and through action, our ADMs feature modest SNR as seen from Fig. 6. For example, for the response of ADM with balanced drop/add action to THz signal with a data rate of 1 Gbps, SNR for through action in grating passband (THz light launched into In port and measured at Through port) and drop (add) action in grating stopband (THz light launched into In (Add) port and measured at Drop (Through) port) is approximately 11 dB and 5 dB, respectively. This is in good agreement with the insertion losses obtained from the spectroscopic measurements, in which the insertion loss is ~10 dB for through action and ~15 dB for drop and add action (see Figs.5(e) and 5(f))

**Supplementary Note 8. Experimental setup to characterize two-wire WBGs**

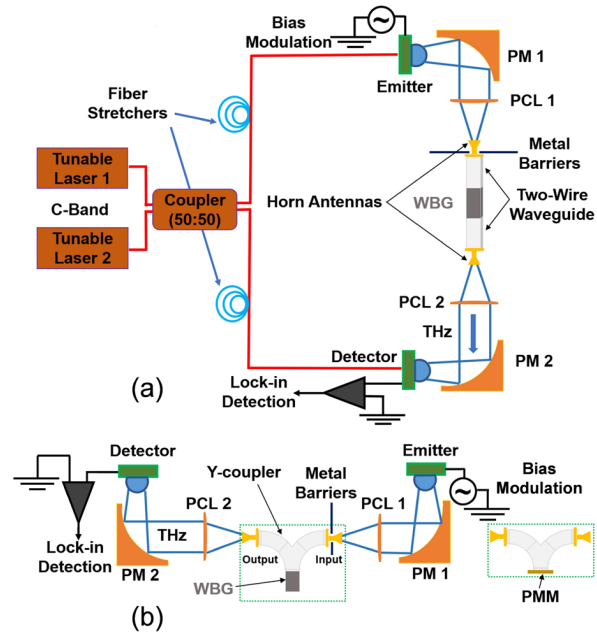

Fig. S7 The continuous-wave (CW) THz spectroscopy system to characterize two-wire WBGs. (a) Experimental setup to measure the relative transmission coefficient of two-wire WBGs. PM: parabolic mirror, PCL: plano convex lens. (b) Experimental setup to measure the relative reflection coefficient of two-wire WBGs. Insert: the THz optical components to replace the ones that has been placed inside the CW-THz spectroscopy system (in green dotted region) for the reference spectrum. PMM: plano metallic mirror.

**Supplementary Note 9. Experimental setup to characterize directional coupler and ADM circuits**

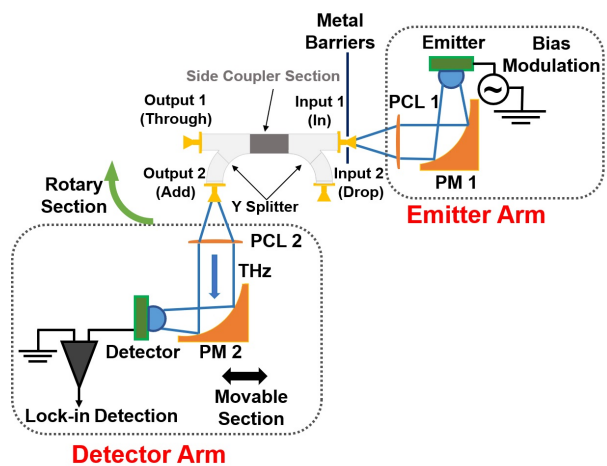

Fig. S8 The CW-THz spectroscopy system to characterize the directional coupler and ADM circuits.

**Supplementary Note 10. Characterization of the two-wire ADM circuits using THz communication system**

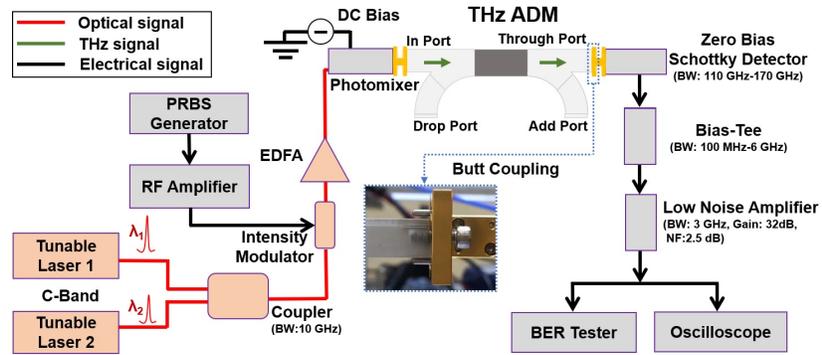

Fig. S9 Schematic of the photonics-based THz communication system to characterize THz ADM circuits. Insert: Butt coupling of the THz ADM circuit and a WR6.5 waveguide connected to the Schottky detector.

## Supplementary Note 11. Numerical simulation of two-wire WBGs

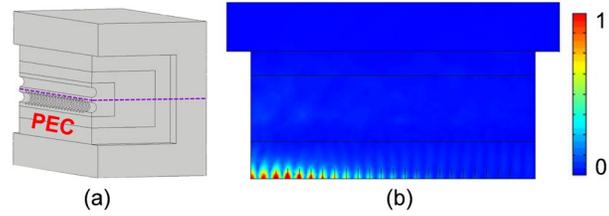

Fig. S10 Numerical simulation of two-wire WBGs. (a) The 3D model built in COMSOL Multiphysics to study WBGs. (b) The electric field distribution of THz light within grating stopband (140 GHz) in a two-wire WBG on the plane noted by the violet dashed line in Fig. S10(a)) when the THz light is launched using port boundary condition at the input facet of the WBG.